\documentclass[fleqn,10pt]{wlscirep}
\usepackage[utf8]{inputenc}
\usepackage[T1]{fontenc}
\title{Demonstration of critical coupling in an active III-nitride microdisk photonic circuit on silicon}

\author[1,2]{Farsane Tabataba-Vakili}
\author[3]{Laetitia Doyennette}
\author[3]{Christelle Brimont}
\author[3]{Thierry Guillet}
\author[4]{St\'{e}phanie Rennesson}
\author[4]{Benjamin Damilano}
\author[4]{Eric Frayssinet}
\author[4]{Jean-Yves Duboz}
\author[1]{Xavier Checoury}
\author[1]{S\'ebastien Sauvage}
\author[1]{Moustafa El Kurdi}
\author[4]{Fabrice Semond}
\author[2]{Bruno Gayral}
\author[4,*]{Philippe Boucaud}
\affil[1]{Centre de Nanosciences et de Nanotechnologies, CNRS, Univ. Paris-Sud, Universit\'{e} Paris-Saclay, F-91120 Palaiseau, France}
\affil[2]{Univ. Grenoble Alpes, CEA, INAC-Pheliqs, 38000 Grenoble, France}
\affil[3]{Laboratoire Charles Coulomb (L2C), Universit\'e de Montpellier, CNRS, Montpellier, France}
\affil[4]{Universit\'{e} C\^{o}te d'Azur, CNRS, CRHEA, F-06560 Valbonne, France}

\affil[*]{philippe.boucaud@crhea.cnrs.fr}


\begin{abstract}
On-chip microlaser sources in the blue constitute an important building block for complex integrated photonic circuits on silicon. We have developed photonic circuits operating in the blue spectral range based on microdisks and bus waveguides in III-nitride on silicon. We report on the interplay between microdisk-waveguide coupling and its optical properties. We observe critical coupling and phase matching, i.e. the most efficient energy transfer scheme, for very short gap sizes and thin waveguides ($g=45$ nm and $w=170$ nm) in the spontaneous emission regime. Whispering gallery mode lasing is demonstrated for a wide range of parameters with a strong dependence of the threshold on the loaded quality factor. We show the dependence and high sensitivity of the output signal on the coupling. Lastly, we observe the impact of processing on the tuning of mode resonances due to the very short coupling distances. Such small footprint on-chip integrated microlasers providing maximum energy transfer into a photonic circuit have important potential applications for visible-light communication and lab-on-chip bio-sensors.
\end{abstract}
\begin{document}

\flushbottom
\maketitle

\thispagestyle{empty}

\section*{Introduction}

Microresonators, like microdisks, microrings or microspheres evanescently coupled to waveguides have been a field of intense study for nearly two decades \cite{Yariv2000}. These devices are basic building blocks for chip-scale integrated photonics and have enabled a large range of applications in telecommunication, optical computing, optical interconnects and sensing. One of the key issues for such a system is the efficient energy transfer between the resonator and waveguide, which is achieved in the so-called critical coupling regime when the rate of energy decay from the cavity to the waveguide  equals the intrinsic rate of energy decay from the uncoupled microresonator. All the energy can be transferred from the the microresonator to the waveguide (or vice versa), by analogy with impedance matching in microwave devices. Experimental demonstrations of critical coupling were reported in several platforms, using silica microspheres coupled to fibers in the near-infrared \cite{Cai2000, Spillane2002}  or in the silicon photonics platform using silicon-on-insulator (SOI) \cite{Bogaerts2012}. Critical coupling is usually controlled through the coupling quality factor that depends on the cavity mode volume, the overlap, interaction length, and the effective mode index mismatch between the waveguide and the microresonator. Critical coupling is obtained by carefully tuning the distance between microresonator and waveguide and by engineering the mode index dispersion in order to reach the phase matching condition.

In the literature, most of the demonstrations of critical coupling have been evidenced by injecting light into a waveguide coupled to a passive microresonator and looking at the transmission. At critical coupling, there is a perfect destructive interference between the transmitted field and the microresonator's internal field coupled to the waveguide, thus leading to quenching of the transmission on resonance. Active microresonators under spontaneous or stimulated emission also constitute a scheme of interest. The cavity can consist of an individual microresonator or a microresonator coupled to waveguides and additional reflectors. Electrically injected microlasers coupled vertically or laterally to silicon bus waveguides have been achieved using indium phosphide (InP) and indium
gallium arsenide (InGaAs) on SOI substrate with emission around $1.5 ~\mu$m \cite{Fang2006, VanCampenhout2008,Liang2009}. Low threshold powers of around 10 mW at room-temperature (RT) under continuous-wave (CW) operation were reported. Using quantum dots in III-arsenide mushroom-type disks and suspended waveguides, shorter-wavelength emission at around 850 nm was demonstrated under optical excitation \cite{Koseki2008}. A review of whispering gallery microcavity lasers can be found in Ref. \cite{He2013}.  In the near-infrared the fabrication constraints are rather relaxed as typical distances between microresonators and waveguides are in the hundreds of nm range.

One very attractive platform for integrated photonics in the visible spectral range is based on III-nitrides on silicon. Nanophotonics utilizing this platform is an emerging field that has certainly raised considerable interest in the past decade. The main advantages of this material system are the possibility to have active monolithically integrated laser sources from the ultra-violet to visible (UV-VIS) spectral range and a large transparency window for energies smaller than 6 eV. There have been numerous demonstrations of microlasers \cite{Tamboli2007,Simeonov2007,Aharonovich2013,Athanasiou2014,Zhang2015,Selles2016} and high quality (Q) factor microresonators using III-nitrides \cite{Simeonov2008,Mexis2011,Rousseau2018}. Several reports have been made on passive microdisks evanescently coupled to bus waveguides in the UV-VIS spectral range \cite{Stegmaier2014,Lu2018,Liu2018}, but only very few demonstrations of active microlaser photonic circuits have been reported \cite{TabatabaVakili2018}, while critical coupling has been observed in passive photonic circuits in the IR \cite{Pernice20122,Jung2013,Bruch2015,Roland2016}. Active photonic circuits using light emitting diodes have been demonstrated, but using inefficiently large device dimensions \cite{Shi2017}. III-nitride on silicon microdisks critically coupled to waveguides pose an important building block for more complex integrated photonic circuits that will greatly benefit the fields of both III-nitride and silicon photonics \cite{Zhou2015}. Potential applications range from visible-light communication \cite{Xie2019}, to optical interconnects \cite{Brubaker2013}, to on-chip quantum optics \cite{Jagsch2018} and lab-on-chip applications, such as bio-sensing \cite{Lia2017} and gene activation \cite{Polstein2012}. 

\begin{figure}[ht]
\centering
\includegraphics[width=1\linewidth]{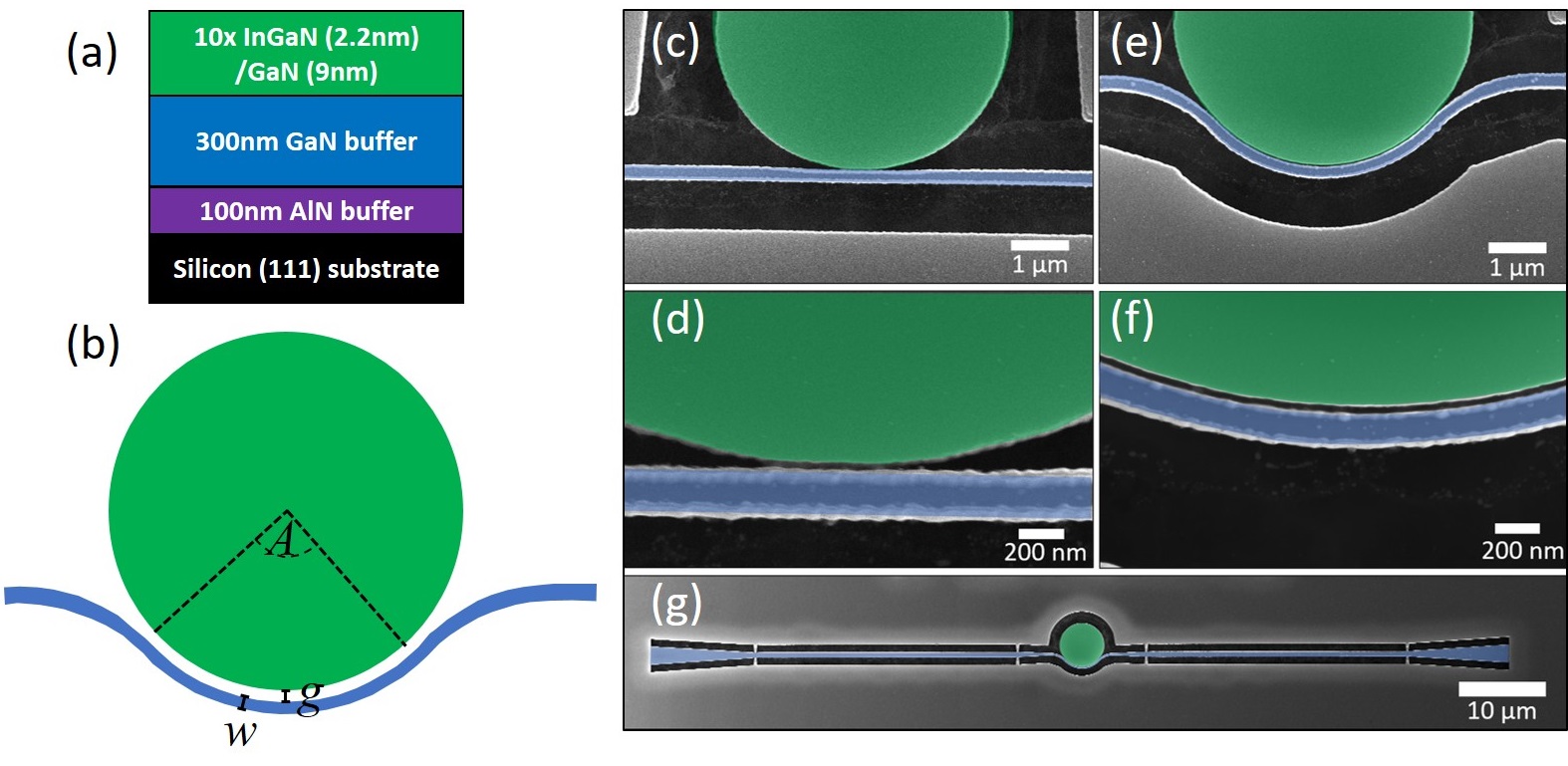}
  \caption{(a) Sketch of the sample heterostructure, (b) Sketch of the device with $A$ - bending angle of the waveguide around the disk, $g$ - gap between the disk and the waveguide, $w$ - waveguide width, (c,d) False colors SEM images of a device with $A = 0^\circ$ and $g = 45 \text{~nm}$, (e,f) SEM images of a device with $A = 90^\circ$ and $g = 45 \text{~nm}$, (g) SEM image of a full device. The green coloring represents the area with indium gallium nitride (InGaN) quantum wells (QWs) (i.e. the microdisk) and the blue areas are etched to the gallium nitride (GaN) buffer layer (i.e. the waveguide).}
  \label{fig:sem}
\end{figure}

In this article, we report on critical coupling in an active III-nitride on silicon microdisk with a bus waveguide and show lasing for a wide range of device parameters in the blue with threshold energy densities as low as $1.2~\text{mJ/cm}^2$ per pulse in the under-coupled regime. To the best of our knowledge this is the shortest wavelength demonstration of critical coupling and consequently shortest coupling gap size reported. The fabrication of such small gaps poses significant technological challenges, as it is approaching the limit of what can be achieved with conventional fabrication means. We report on a robust suspended topology with a bent waveguide design \cite{TabatabaVakili2018} to facilitate reaching critical coupling\cite{Bogaerts2012, Bruch2018}. We demonstrate lasing over a wide range of gap sizes and waveguide bending angles, observing a strong dependence of the lasing threshold on gap size and loaded quality factor ($Q_{\text{loaded}}$). We determine the intrinsic Q factor ($Q_{\text{int}}$) to be 4700 and observe a reduction in $Q_{\text{loaded}}$ by a factor 2 for a gap of $45~\text{nm}$, indicating critical coupling. 
We model the output power as a function of the coupling Q factor $Q_C$ and determine the maximum as a function of threshold power and pump power. This maximum is attained for values near the critical coupling point of the spontaneous emission regime. A shift in mode position at small gap size is observed due to a reduction in disk diameter caused by a proximity effect of the waveguide during e-beam lithography. The control of this shift is a key feature for the fine-tuning of the resonator modes.

\section*{Results and Discussion}

\subsection*{Critical coupling in the spontaneous emission regime}

\begin{figure}[htbp]
\centering
\includegraphics[width=0.7\linewidth]{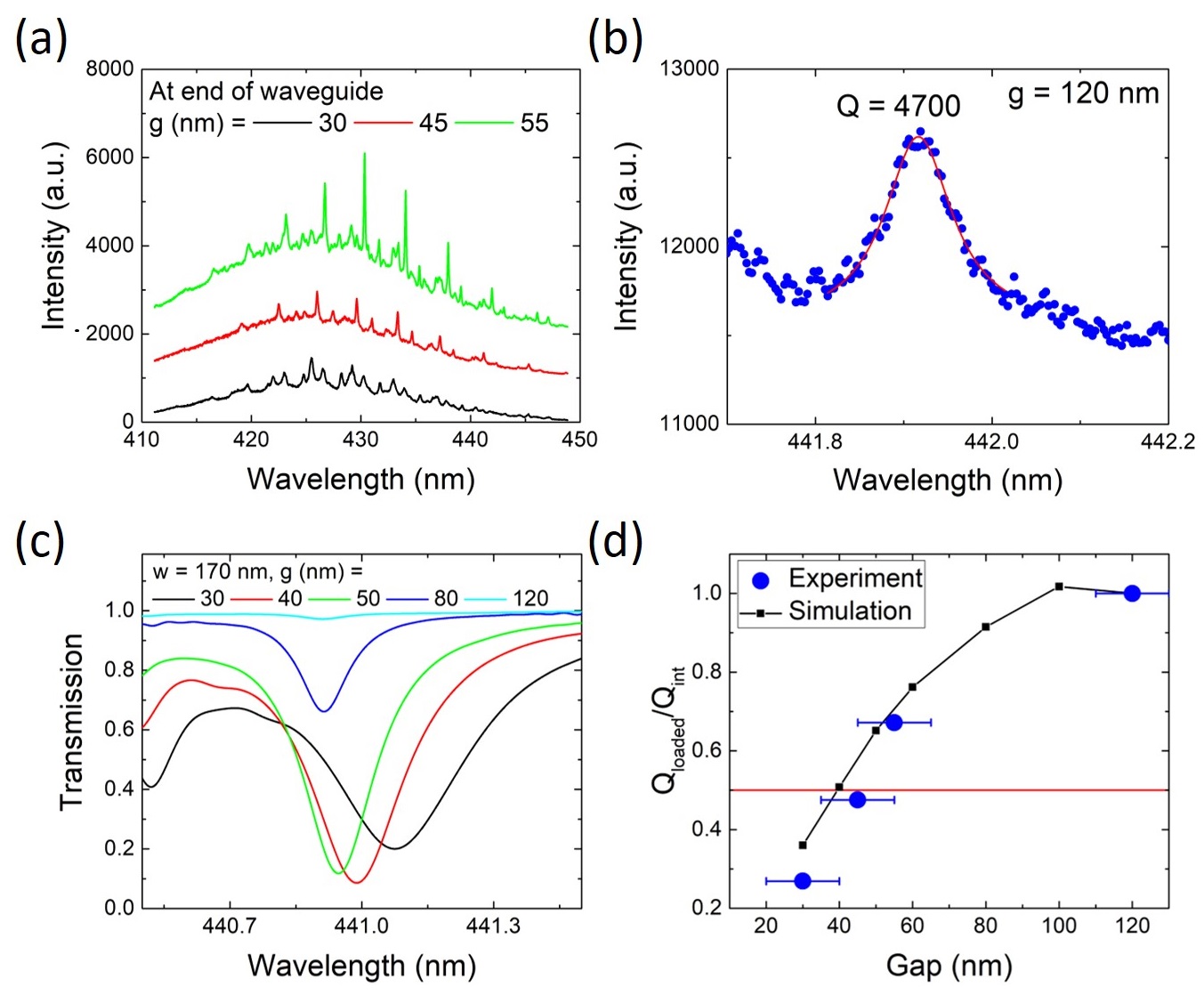}
  \caption{Critical coupling of $5~\mu$m diameter disks: (a) CW $\mu$-PL spectra taken at the end of the waveguide for devices with a waveguide bending angle $A= 90^\circ$ and $g = 30$ to 55 nm. (b) CW $\mu$-PL spectrum measured above the disk for a device with $g=120$ nm measured using a 3600 grooves/mm grating. The mode is fitted with a Lorentzian, giving $Q_{\text{int}}=4700$. (c) FDTD simulation of the radiative flux transmitted through the end of the waveguide of devices with $g= 30$ to 120 nm. (d) Comparison of experimental and simulated results of $Q_{\text{loaded}}/Q_{\text{int}}$ vs. gap for the 442 nm mode. The red horizontal line indicates critical coupling.} 
  \label{fig:cw}
\end{figure}

 The heterostructure of the sample investigated in this study is shown in Fig. \ref{fig:sem} (a) and described in detail in the Methods section.
We fabricated $5~\mu \text{m}$ diameter microdisks with bent bus waveguides with bending angles between $A= 0^\circ$ and $90^\circ$, a waveguide width of nominally $w=135$ nm (measured: $w=170$ nm), and nominal gap sizes between the disk and the waveguide of $g=40$ to 120 nm (measured: $g=30$ to 120 nm).  A sketch of a device is shown in Fig. \ref{fig:sem} (b) highlighting the parameters $A$, $g$, and $w$. False color scanning electron microscopy (SEM) images of devices with $g=45$ nm are shown in Fig. \ref{fig:sem} (c,d) for $A= 0^\circ$ and (e,f) for $A= 90^\circ$. In the zoom-ins in Fig. \ref{fig:sem} (d) and (f) it is clearly visible that the small gaps are fully open. We will from now on refer only to the measured gap sizes. See Fig. S1 in the supplementary information for how these values were determined. Details on the processing can be found in the Methods section.

In order to determine the contribution of the disk-waveguide coupling to the Q factor, we performed RT CW micro-photoluminescence ($\mu$-PL) measurements on devices with $A= 90^\circ$ using a laser emitting at 355 nm and a 20x microscope objective. The microdisk is excited with an $8~\mu$m diameter spot size and the emission is collected from the top, using a spectrometer and a charge-coupled device (CCD) as the detector. By defining areas on the CCD the position of the emission can be discerned (i.e. above the disk or at the end of the waveguide). Spectra of devices with $g=30$ to 55 nm, measured at the end of the waveguide, are shown in Fig. \ref{fig:cw} (a). First-order radial whispering gallery modes (WGMs) are clearly visible and the azimuthal mode orders are identified in Fig. S2 and S3 in the supplementary information through finite-difference time-domain (FDTD) simulations. 
$Q_\text{loaded}$ is given by

\begin{equation}
    \frac{1}{Q_\text{loaded}} = \frac{1}{Q_\text{int}} + \frac{1}{Q_C(g)} + \frac{1}{Q_\text{abs,QW}(\lambda)}, \label{eq:Q}
\end{equation}

where $Q_C(g)$ is the coupling Q factor as a function of the gap $g$ and $Q_\text{abs,QW}(\lambda)$ is the absorption Q factor of the QWs as a function of $\lambda$. We define $Q_\text{int}$ as the Q factor of an uncoupled microdisk excluding quantum well absorption. This quantity takes factors like roughness and residual absorption into account. $Q_\text{int}$ is determined by investigating a device with $g=120$ nm, where the coupling to the bus waveguide becomes negligible ($Q_{\text{loaded}}\approx Q_{\text{int}}$ in the under-coupled regime). Fig. \ref{fig:cw} (b) shows the spectrum of a mode with $Q_\text{int} \approx  Q_\text{loaded} =4700$, given by $Q=\lambda/\Delta \lambda$, where $\lambda$ is the resonant wavelength and $\Delta \lambda$ is the full-width at half maximum (FWHM) of the resonance, determined by a Lorentzian fit and measured at the disk for a device with $g=120$ nm, where the coupling is very weak, and in the low energy tail of the spectrum, where the contribution of the QW absorption is negligible, and using a 3600 grooves/mm grating. $Q_\text{int}=4700$ is state of the art for III-nitrides in the blue spectral range. Slightly higher values have been achieved using oxygen passivisation \cite{Rousseau2018}. The main limitation for the Q factor at short wavelength are scattering losses that scale with $\lambda^{-4}$ due to sidewall roughness. At $\lambda=420$ nm we estimate $Q_{abs,QW}=5000$ in analogy to Ref. \cite{TabatabaVakili2017}. At longer wavelength, the quantum well absorption vanishes as well as its contribution to $Q_\text{loaded}$. In Fig. \ref{fig:cw} (a), the linewidth of the whispering gallery modes does not change for $\lambda > 430 ~\text{nm}$, since the absorption is negligible. As will be shown later, $Q_C$ is larger than $10^5$ for a gap size of 120 nm. At critical coupling $Q_\text{loaded}= 1/2 \cdot Q_\text{int}$. The occurrence of critical coupling is associated with a maximum in energy transfer from the microdisk to the waveguide as a function of the gap. FDTD simulations show the sensitive dependence of the transmitted radiative flux for a mode at $441$ nm in Fig. \ref{fig:cw} (c) for gaps between 30 and 120 nm for a $5~\mu$m diameter disk. Experimental and simulated $Q_\text{loaded}$ factors are plotted as a function of gap in Fig. \ref{fig:cw} (d). The experimental values are determined for the 442 nm mode where absorption is negligible, since this is the low-energy end of the QW emission/absorption spectrum. Critical coupling is attained at $g \approx 40-45$ nm for both experiment and simulation. Phase matching of the modes in the disk and waveguide is a necessary condition for critical coupling. It is achieved by fine-tuning $w$ and $g$ to get the mode profiles in the waveguide and disk to match. Figure S3 in the supplementary information depicts FDTD simulations of the $H_z$ field for $5~\mu\text{m}$ disks with  $A= 90^\circ$ and $g=30$ to 50 nm. Good phase matching can be observed for both $g=40$ and 50 nm.

\begin{figure}[htbp]
\centering
\includegraphics[width=0.5\linewidth]{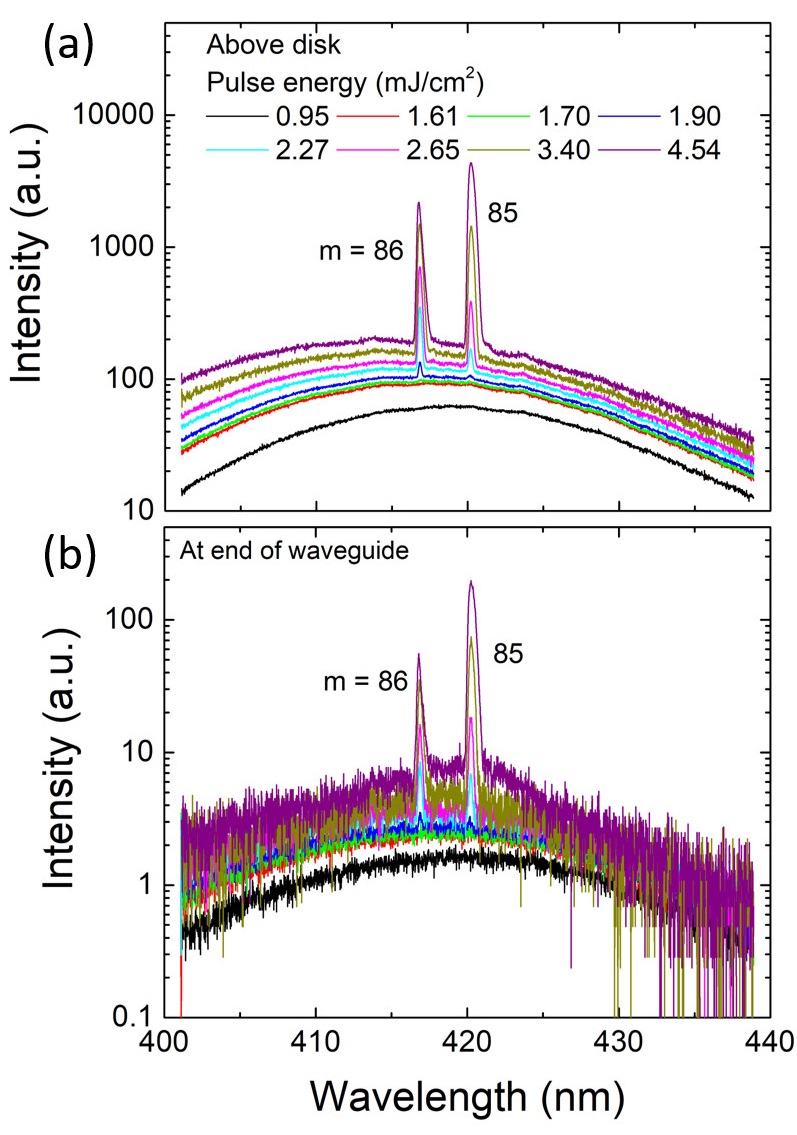}
  \caption{Lasing at the critical coupling gap size: pulse energy dependent spectra measured (a) above the disk, (b) at the end of the waveguide for a device with $A= 90^\circ$ and $g = 45\text{~nm}$. The azimuthal numbers of the first-order radial modes are $m=86$ and 85. The threshold energy density of the $m=86$ mode is  $1.7~\text{mJ/cm}^2$ per pulse.}
  \label{fig:lase}
\end{figure}

\subsection*{Dependence of lasing threshold and output signal on Q factor and coupling distance}

\begin{figure}[htbp]
\centering
\includegraphics[width=0.8\linewidth]{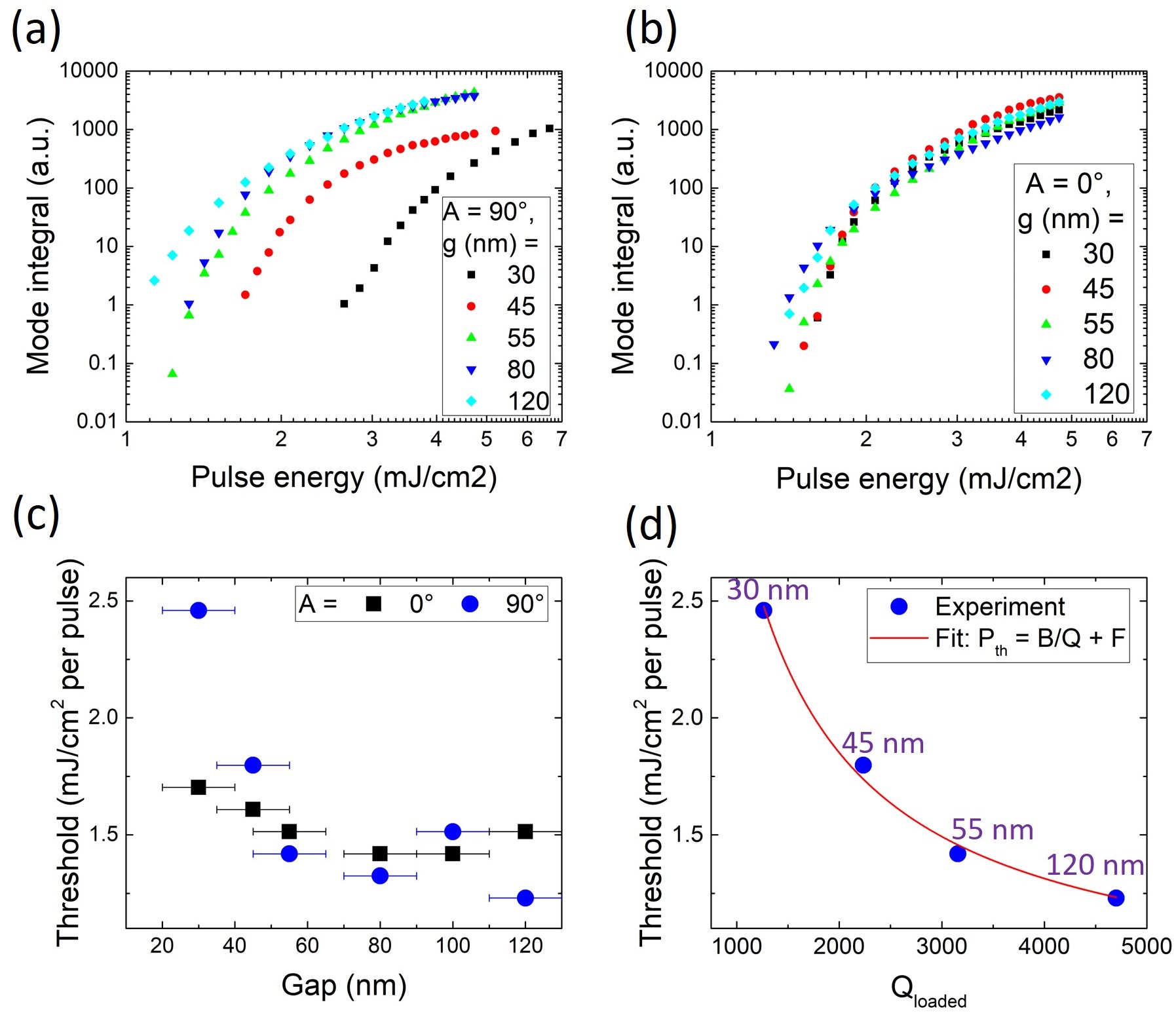}
  \caption{Mode integral over pulse energy for different gap sizes for a 5 $\mu$m disk for (a) $A= 90^\circ$ and (b) $A= 0^\circ$. (c) Threshold vs. gap for $A= 90^\circ$ and $A= 0^\circ$. (d) Threshold vs. $Q_{\text{loaded}}$ for $A= 90^\circ$ using the $Q_{\text{loaded}}$ values from CW excitation in Fig. \ref{fig:cw} (a,b) and fitted with equation (\ref{eq:Pth}).}
  \label{fig:thresh}
\end{figure}

Using a pulsed 355 nm laser with 7 kHz repetition rate, 4 ns pulse width and a 20x microscope objective in a standard $\mu$-PL setup, lasing was demonstrated at RT for all investigated values of $A$ and $g$. Figure \ref{fig:lase} shows pulse energy dependent spectra of a device with $A= 90^\circ$ and $g=45$ nm, corresponding to the critical coupling parameters under CW excitation discussed in Fig. \ref{fig:cw} (d). The spectra in Fig. \ref{fig:lase} (a) are taken using  top-collection above the disk and in Fig. \ref{fig:lase} (b) at the end of the waveguide. Two first-order radial modes at 416 and 420 nm are lasing. Their azimuthal numbers are determined to be $m=86$ and 85 by comparison with FDTD simulations (see Figs. S2 and S3 in the supplementary information). The threshold energy density is determined as the value where the mode starts to become clearly visible in the spectrum. For the $m=86$ mode $P_{th}=1.7~\text{mJ/cm}^2$ per pulse. The other below-threshold modes are not visible due to the pulsed excitation with top-collection, a configuration that does not allow for easy detection of modes below threshold \cite{Selles2016}. Similar lasing spectra for $g= 30$ nm and $g= 55$ nm are shown in Fig. S6 in the supplementary information.

The lasing mode integral vs. pulse energy is shown in Fig. \ref{fig:thresh}(a) and (b) for devices with different gaps and for $A= 90^\circ$ and $0^\circ$, respectively. Fig. \ref{fig:thresh} (c) summarizes the thresholds for both angles. As previously observed in Fig. \ref{fig:cw} (a-c) in the spontaneous emission regime, the impact of the gap size on the lasing behavior is very strong for values near the spontaneous emission critical coupling point. A larger effect of the gap size on the threshold is observed for $A= 90^\circ$ than for $0^\circ$. A geometry with $A= 90^\circ$ allows for critical coupling at a larger distance than the straight waveguide configuration, due to the increased coupling length \cite{Hu2008}. For $A= 0^\circ$, the critical coupling distance is shorter and was not reached experimentally. The lowest threshold of $1.2~\text{mJ/cm}^2$ per pulse is observed for  $A= 90^\circ$ and $g=120$ nm (see Fig. S7 (a) in the supplementary information). Linewidth narrowing of more than a factor of two is observed when approaching the threshold (Fig. S7 (b)).  Figure \ref{fig:thresh} (d) shows the lasing threshold as a function of $Q_{\text{loaded}}$, using the below-threshold CW $Q_{\text{loaded}}$ values from Fig. \ref{fig:cw} and the threshold values from Fig. \ref{fig:thresh} (c). It is interesting to quantitatively analyze the microlaser threshold and the collected laser emission as a function of the coupling strength. The threshold is given as \cite{Baba2003,Gargas2009}

\begin{align}
    P_{th} & \propto G_{th} = \frac{2 \pi n_g}{\Gamma\lambda Q_{\text{loaded}}} +G_{tr},\\
        P_{th} & = B\cdot\big(\frac{1}{Q_{\text{int}}}+\frac{1}{Q_C}\big)+F, \label{eq:Pth}
\end{align}

with $G_{th}$ the threshold gain, $n_g$ the group index in the material, $\Gamma$ the energy confinement factor, $\lambda$ the resonant wavelength, $G_{tr}$ the gain needed to achieve transparency, and $B=2070$ and $F=0.79$ fit parameters. The red curve in Fig. \ref{fig:thresh} (d) is given by equation (\ref{eq:Pth}) and matches well with the experimental data, showing that the laser threshold can be controlled by tuning $Q_{\text{loaded}}$ by adjusting the gap size.

\begin{figure}[htbp]
\centering
\includegraphics[width=0.5\linewidth]{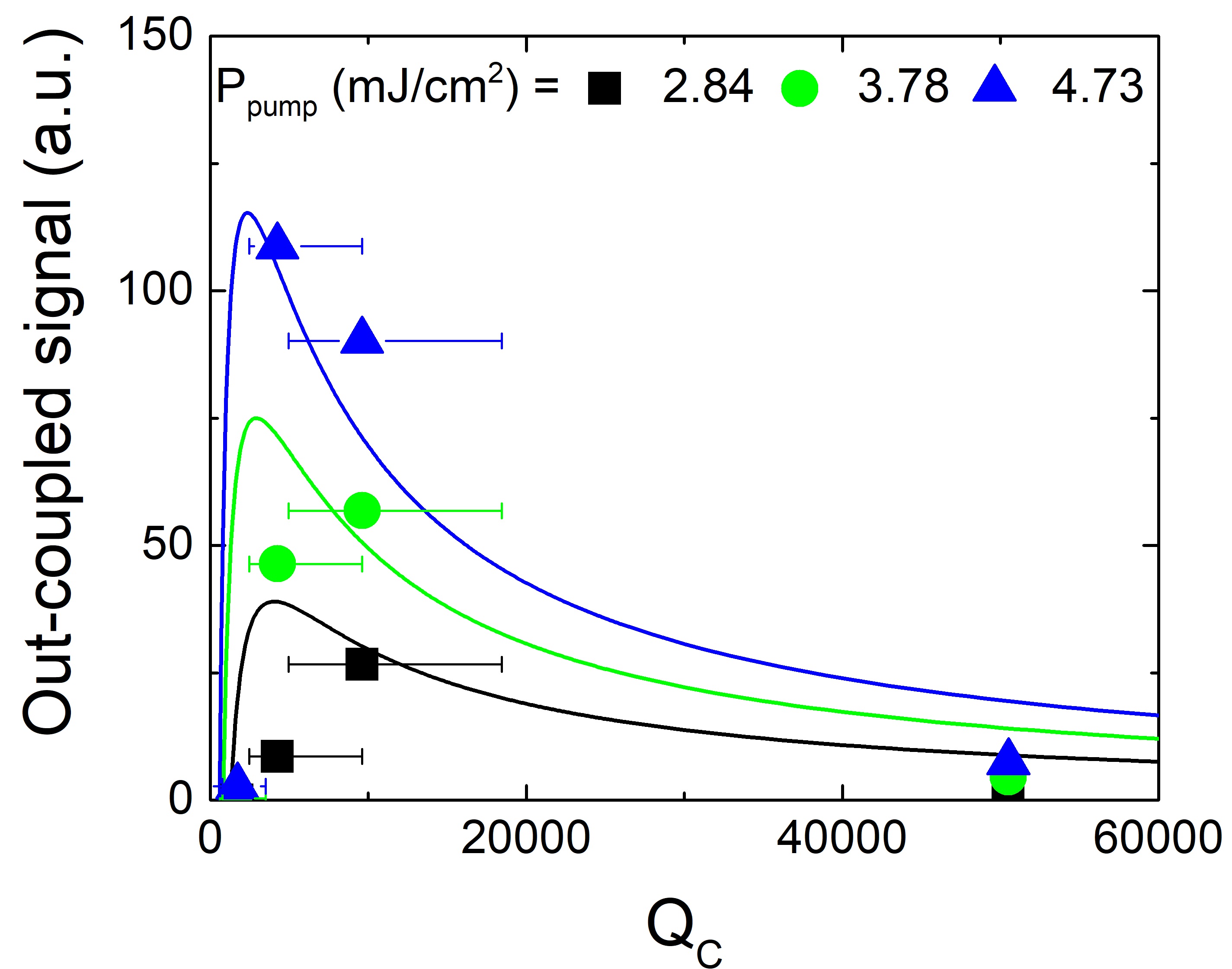}
  \caption{Out-coupled signal integrated over one mode as a function of $Q_C$ for $A= 90^\circ$ measured at the end of the waveguide (symbols) and fitted with equation (\ref{eq:Pout}).}
  \label{fig:out}
\end{figure}

As seen in Fig. \ref{fig:thresh}, the threshold depends on the coupling gap between the microdisk and the bus waveguide as well as on the $Q_{\text{loaded}}$. The energy extracted at the end of the waveguide depends also on $Q_C$ and there is an optimal $Q_C$ providing maximum extracted energy. In the case of standard ridge lasers the optimization of the output power is well-known and the maximum out-coupled power depends on the mirror reflectivity $R$ \cite{Rosencher2002}. For high-power lasers a small $R$ is chosen, providing large out-coupled power but a higher threshold, while for low-power lasers $R$ needs to be large and the out-coupled power and threshold are small. The cases of microsphere, microdisk or microring lasers coupled with bus waveguides have been thoroughly investigated as well. In most cases, the pump energy was injected through the bus waveguide \cite{Cai2000,Min2004,Bogaerts2012,Rasoloniaina2014}, which results in the laser characteristics being doubly dependent on the coupling to waveguides for both pump injection and laser emission. In the case investigated here, population inversion is achieved through an external pump that is not linked with the bus waveguide. For this type of configuration, which is also relevant for an electrical injection scheme, the know-how to achieve optimum out-coupled power is essential. This optimum depends on the pump energy as well as the microdisk to waveguide coupling. In the spontaneous emission regime the energy transfer is most efficient at critical coupling. However, in the lasing regime the gap distance providing maximum energy transfer does not necessarily coincide with the spontaneous emission critical coupling. It is thus necessary to investigate the dependence of the output signal on the pump energy and coupling. Figure \ref{fig:out} shows the output signal measured at the end of the waveguide and integrated over one mode as a function of $Q_C$ for different pump powers $P_{\text{pump}}$ for devices with $A= 90^\circ$. This signal is proportional to the out-coupled power $P_{\text{out}}$ and is integrated over the mode that shows lasing first. The experimental data points are fitted using \cite{Siegman1986,Yariv1989,Min2004,He2013}

\begin{equation}
P_{\text{out}} = C \cdot \frac{\frac{1}{Q_C}}{\big(\frac{1}{Q_{\text{int}}}+\frac{1}{Q_C}\big)}\cdot (P_{\text{pump}}-P_{th}),\label{eq:Pout}
\end{equation}

where $C=66$ and $P_{th}$ is given by equation (\ref{eq:Pth}) and the fit in Fig. \ref{fig:thresh} (d). This equation reflects the external quantum efficiency given by the ratio between coupling losses and global losses and the threshold condition of the laser. The $Q_C$ values were determined using equation (\ref{eq:Q}), where $Q_{\text{int}}=4700$ is taken from Fig. \ref{fig:cw} (b), and $Q_{\text{loaded}}$ values are taken from Fig. \ref{fig:cw} (d). The error bars are determined via an exponential fit between $Q_C$ and gap and the $\pm 10$ nm error assumed for the gaps. The maximum of $P_{\text{out}}$, representing maximum energy transfer, shifts towards smaller $Q_C$ with increasing $P_{\text{pump}}$. For large values of $Q_C$ the power quickly drops towards zero. Experiment and calculation match rather well. The quality of the fit is limited by the accuracy of the gap values and by the limited number of experimental data points. Figure S5 in the supplementary information shows the maximum coupling gap over $P_{\text{pump}}$ calculated using equation (S3). The maximum energy transfer under lasing conditions does not occur for the same $Q_C$ and gap as critical coupling ($Q_C=Q_{\text{int}}=4700$) in the spontaneous emission regime, but at slightly smaller values in the investigated range of $P_{\text{pump}}$ (i.e. $Q_C=2400$ for $P_{\text{pump}}=4.73~\text{mJ/cm}^2$ per pulse), and the $P_{\text{pump}}$ dependence allows for fine-tuning of the coupling efficiency for a given $Q_C$ and gap size.

\subsection*{Proximity effect}

\begin{figure}[htbp]
\centering
\includegraphics[width=0.8\linewidth]{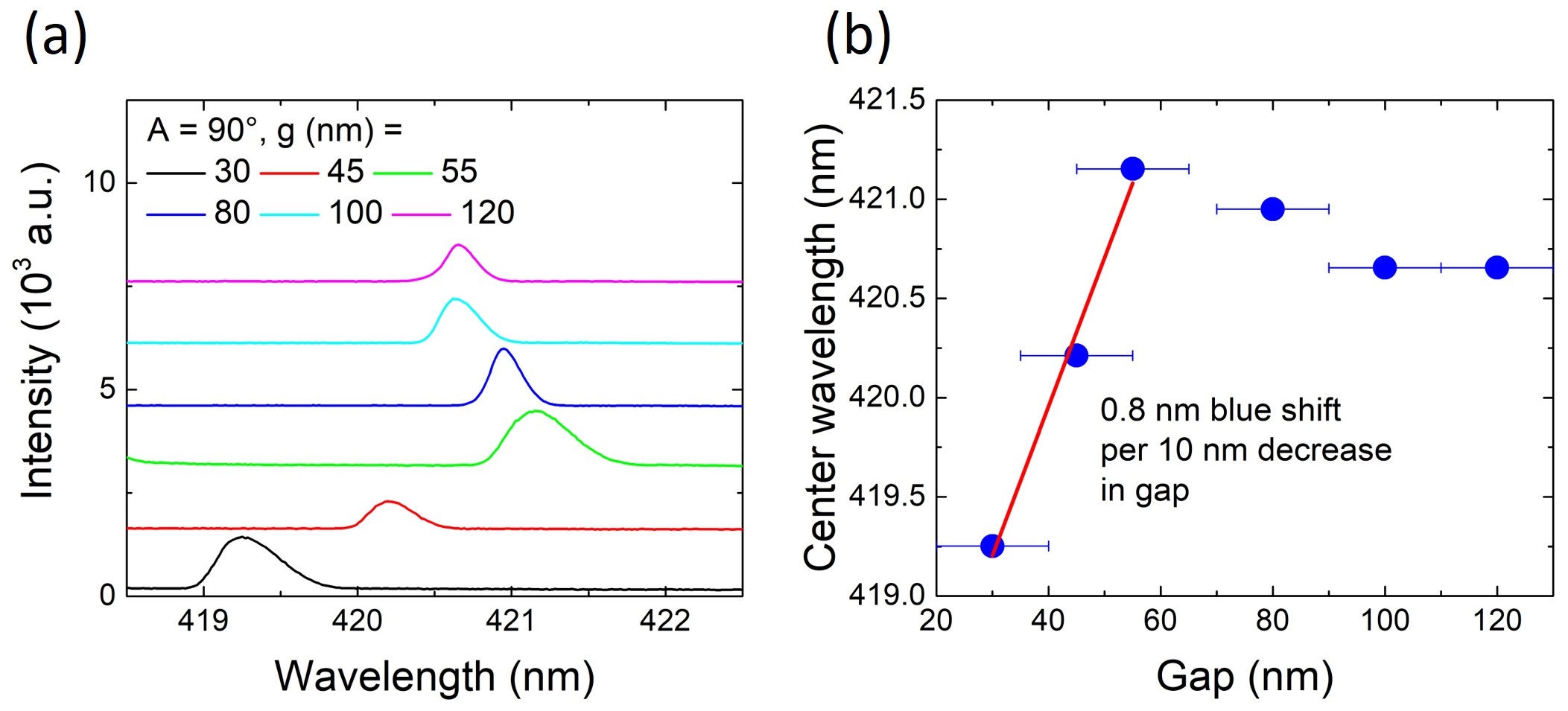}
  \caption{(a) Pulsed lasing spectra measured at the disk for devices with $A= 90^\circ$ and different gap sizes. b) Peak wavelength as a function of gap for the spectra in (a).}
  \label{fig:shift}
\end{figure}

As the coupling distances are in the tens of nm range for blue emitters, the proximity between waveguide and microresonator is an issue for monolithic integrated circuits as opposed to systems like a microsphere coupled to an external tapered fiber. We have thus investigated the impact of this proximity on the spectral mode resonance. Figure \ref{fig:shift} (a) shows the 421 nm mode of devices with $A= 90^\circ$ for different gap sizes. With decreasing gap size the mode position red shifts and subsequently blue shifts, which is plotted in Fig. \ref{fig:shift} (b). The red shift is expected because of the change in the effective index seen by the whispering gallery mode as the coupling distance decreases. A 0.5 nm red-shift is observed going from a 120 to a 55 nm gap. A small red-shift is also observed in FDTD simulations. For small gaps between 55 and 30 nm a 0.8 nm blue-shift per 10 nm decrease in gap size is observed. This blue-shift is caused by a reduction in disk diameter during the fabrication process due to the proximity of the waveguide \cite{Bogaerts2012}. The FDTD simulations in Fig. S4 in the supplementary information show that a 10 nm decrease in diameter for a $5~\mu\text{m}$ disk (0.2\%) causes a $0.8 ~\text{nm}$ blue shift. The reduction in disk diameter is thus roughly directly proportional to the decrease in gap size and in agreement with the experimental observation. The effect of the bending angle on mode position is investigated in Fig. S8 in the supplementary information. The proximity effect is less pronounced for small angles due to a reduced coupling length. These measurements highlight the strong sensitivity of the microresonator system on the coupling distances. A short distance, i.e in the range of tens of nm, is mandatory for efficient coupling. Meanwhile, we reach a regime where the distances set new challenges for fabrication as the gap needs to be open and where the distances have an impact on the microresonator resonances through a loading effect. This effect needs to be taken into account for the design of integrated photonic circuits in the blue spectral range.

\section*{Conclusion}

In conclusion, we have demonstrated critical coupling and lasing in an active III-nitride on silicon nanophotonic circuit utilizing a microdisk and a bent bus waveguide in the blue spectral range. This study was meant to identify the salient features associated with critical coupling and lasing in the blue. Low threshold energies of $1.7~\text{mJ/cm}^2$ per pulse at the critical coupling gap and $1.2~\text{mJ/cm}^2$ per pulse in the under-coupled regime have been observed. Large $Q_{\text{int}}$ factors of 4700 were measured in the under-coupled regime and a reduction in $Q_{\text{loaded}}$ of a factor of 2 was observed at critical coupling for a gap of 45 nm. A strong dependence of $P_{th}$ on $Q_{\text{loaded}}$ was observed that perfectly matches the expectation. $P_{\text{out}}$ was investigated as a function of $Q_C$, and we analyzed the dependence on $P_{\text{pump}}$, finding a good agreement between the experimental data and analytical formula. A proximity effect of the waveguide on the disk was shown to cause a shift in mode position, emphasizing the impact of an optimized design and fabrication on the intrinsic properties of the microresonator.

More complex photonic circuits using active microlasers operating near critical coupling parameters can be used for visible-light communication \cite{Islim2017} and lab-on-chip applications \cite{Estevez2012}. An important step will be to integrate electrical injection into this nanophotonic platform. We have recently developed a compatible technology \cite{TabatabaVakili2019}. Going towards the UV spectral range can also be interesting, for example for germicidal irradiation, however, smaller gap sizes will be necessary for critical coupling, which will be more difficult to master experimentally. Using a bonding process, III-nitride thin films can also be transferred onto SiO$_2$-on-Si substrates, which would eliminate the necessity of underetching, allowing for the fabrication of ring resonators, for example.

\section*{Methods}
\subsection*{Sample growth}
The sample investigated in this study was grown by molecular beam epitaxy (MBE) on silicon (111) substrate. First a 100 nm aluminum nitride (AlN) buffer layer was grown, followed by $300 ~\text{nm}$ GaN and 10 pairs of 2.2 nm In$_{0.12}$Ga$_{0.88}$N / 9 nm GaN QWs, emitting at 430 nm. A sketch of the heterostructure is shown in Fig. \ref{fig:sem} (a). When comparing this sample to the one investigated in our previous work \cite{TabatabaVakili2018}, we observe a factor 4 higher PL intensity from the as-grown sample and a factor 2 improvement in the microdisk $Q_{\text{int}}$, likely due to the fact that there is no silicon-doping in this sample, which improves the material quality.

\subsection*{Sample processing}
We use a process similar to the one described previously \cite{TabatabaVakili2018}. Two layers of e-beam lithography using UVIII resist, reactive ion etching (RIE) of a plasma enhanced chemical vapor deposited (PECVD) silicon dioxide (SiO$_2$) hard mask, and inductively coupled plasma (ICP) etching of the III-nitride using chlorine (Cl$_2$) and boron trichloride (BCl$_3$) gases are used. In the first level the microdisk and bus waveguide are defined and the surrounding area is etched to the Si substrate. The end of the waveguide is tapered from 500 nm to $2~\mu$m width. In the second level the waveguide is etched to the GaN buffer layer to avoid re-absorption of the emission, creating a 120 nm high edge at the end of the waveguide, which allows for light extraction by scattering. Further optimization of light extraction requires processing of grating couplers at the waveguide extremities \cite{TabatabaVakili2018}. As a final step, the devices are underetched using xenon difluoride (XeF$_2$) gas, leaving the microdisk standing on a central pedestal and the waveguide suspended in air, held by tethers. The device length is $100~\mu\text{m}$, as can be seen in Fig. \ref{fig:sem} (g).

\bibliography{crit}

\section*{Acknowledgements}

This work was supported by Agence Nationale de la Recherche under
MILAGAN convention (ANR-17-CE08-0043-02). This work was also partly
supported by the RENATECH network. We acknowledge support by a public
grant overseen by the French National Research Agency (ANR) as part
of the \textquotedblleft Investissements d\textquoteright Avenir\textquotedblright{}
program: Labex GANEX (Grant No. ANR-11-LABX-0014) and Labex NanoSaclay
(reference: ANR-10-LABX-0035).


\newpage

\begin{center}
  \textbf{\large Supporting Information for \\
Demonstration of critical coupling in an active III-nitride microdisk photonic circuit on silicon}\\[.2cm]
  Farsane Tabataba-Vakili,$^{1,2}$ Laetitia Doyennette,$^{3}$ Christelle Brimont,$^3$ Thierry Guillet,$^3$ St\'{e}phanie Rennesson,$^4$ Benjamin Damilano,$^4$ Eric Frayssinet,$^4$  Jean-Yves Duboz,$^4$  Xavier Checoury,$^1$ S\'ebastien Sauvage,$^1$  Moustafa El Kurdi,$^1$ Fabrice Semond,$^4$ Bruno Gayral$^2$ and Philippe Boucaud$^{4,*}$   \\[.1cm]
   {\itshape ${}^1$Centre de Nanosciences et de Nanotechnologies, CNRS, Univ. Paris-Sud, Universit\'{e} Paris-Saclay, F-91120 Palaiseau, France\\
  ${}^2$Univ. Grenoble Alpes, CEA, INAC-Pheliqs, 38000 Grenoble, France\\
  ${}^3$Laboratoire Charles Coulomb (L2C), Universit\'e de Montpellier, CNRS, Montpellier, France\\
  ${}^4$Universit\'{e} C\^{o}te d'Azur, CNRS, CRHEA, F-06560 Valbonne, France\\}
  ${}^*$Electronic address: philippe.boucaud@crhea.cnrs.fr\\
\end{center}

\setcounter{equation}{0}
\setcounter{figure}{0}
\setcounter{table}{0}
\setcounter{page}{1}
\renewcommand{\theequation}{S\arabic{equation}}
\renewcommand{\thefigure}{S\arabic{figure}}

\maketitle

\begin{abstract}
Supporting information for Demonstration of critical coupling in an active III-nitride microdisk photonic circuit on silicon. Device characterization, FDTD simulations, maximum output power, and lasing and mode shift are discussed in more detail.
\end{abstract}

\flushbottom
\maketitle

\renewcommand{\thepage}{S\arabic{page}} 
\renewcommand{\thesection}{S\arabic{section}}  
\renewcommand{\thetable}{S\arabic{table}}  
\renewcommand{\thefigure}{S\arabic{figure}}

\thispagestyle{empty}

\section*{Device characterization}

\begin{figure}[htbp]
\centering
\includegraphics[width=0.8\linewidth]{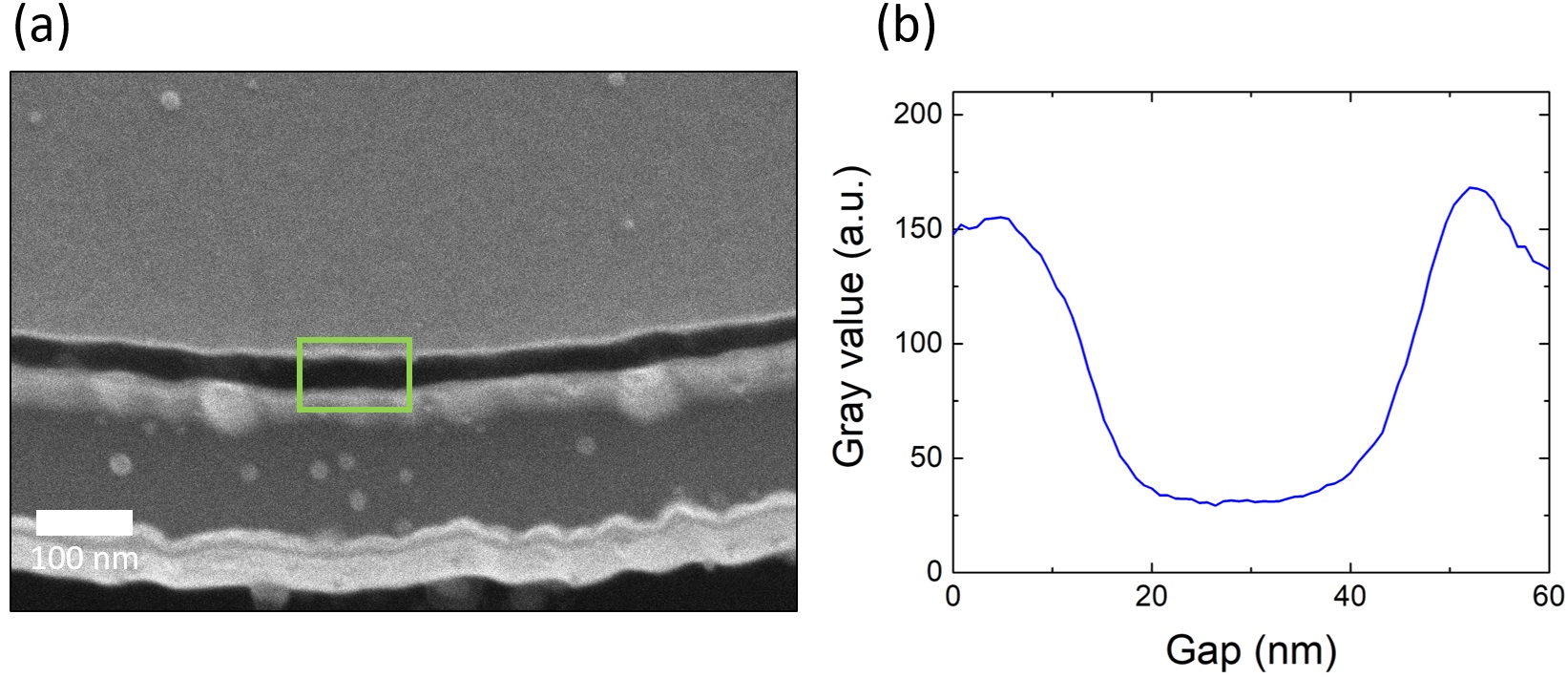}
  \caption{(a) SEM image of a nominally 50 nm gap and (b) gray scale profile along the vertical direction and integrated along the horizontal direction of the rectangle in (a), giving a gap of 45 nm at the top.}
  \label{fig:gap}
\end{figure}

Figure \ref{fig:gap} (a) shows an SEM image of a device with a gap of nominally 50 nm. The gray scale profile in Fig. \ref{fig:gap} (b) is measured along the vertical direction and integrated along the horizontal direction of the rectangle in Fig. \ref{fig:gap} (a). The measurement gives a gap of 45 nm at the top. With this method we determined the gaps provided in the main text. 

\section*{FDTD simulations}

\begin{figure}[htbp]
\centering
\includegraphics[width=0.6\linewidth]{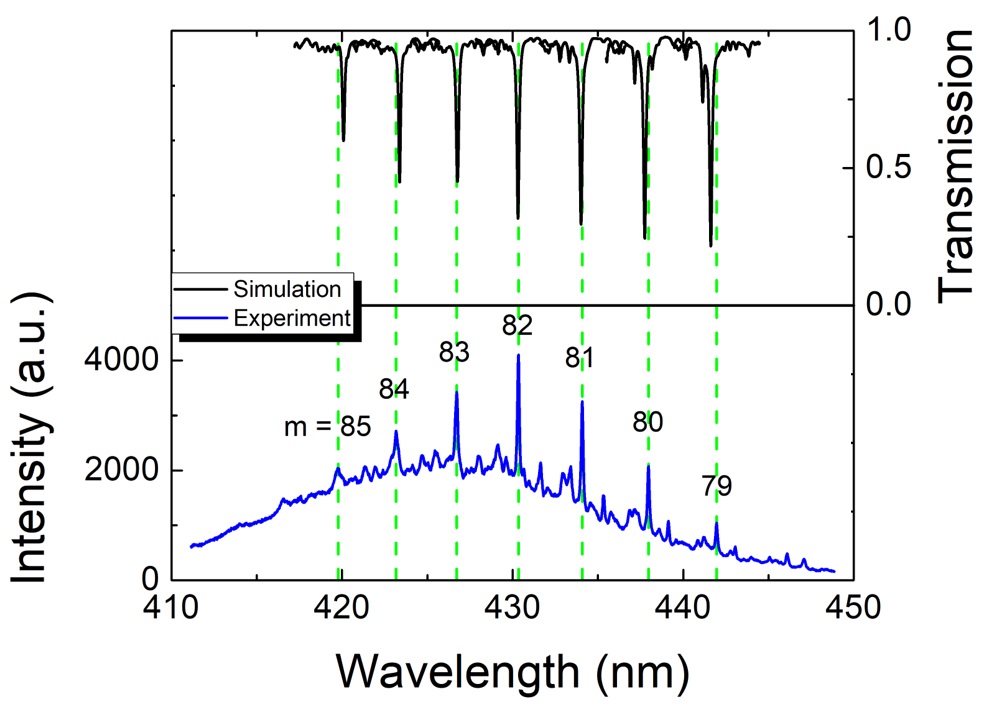}
  \caption{Simulated (top) and experimental (bottom) spectra of a device with $A=90^\circ$ and $g=50$ nm (simulated) and 45 nm (experimental). The azimuthal orders of the first-order radial modes are $m=85$ to 79.}
  \label{fig:spec}
\end{figure}

\begin{figure}[htbp]
\centering
\includegraphics[width=0.8\linewidth]{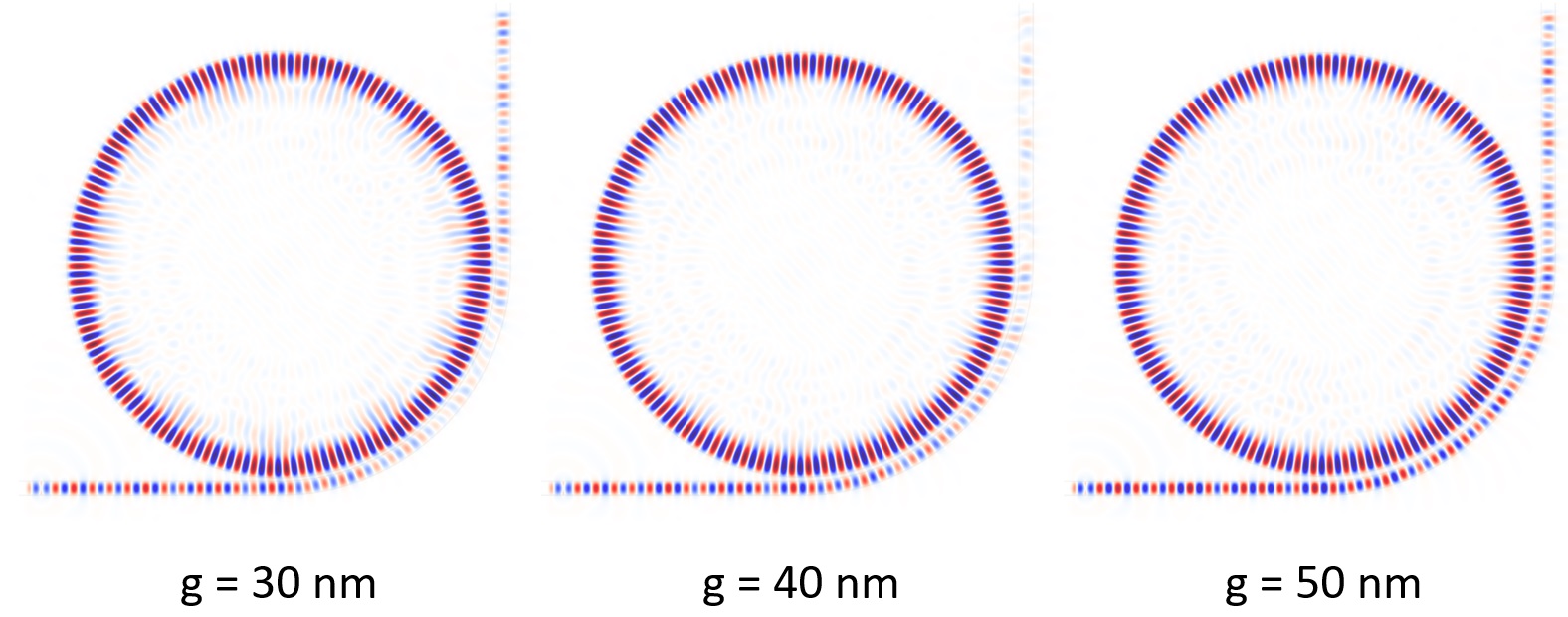}
  \caption{FDTD simulations of the $H_z$ field for devices with $5~\mu \text{m}$ diameter, $w=170$ nm, and $g=30$ nm to 50 nm.}
  \label{fig:sim5}
\end{figure}

\begin{figure}[htbp]
\centering
\includegraphics[width=0.6\linewidth]{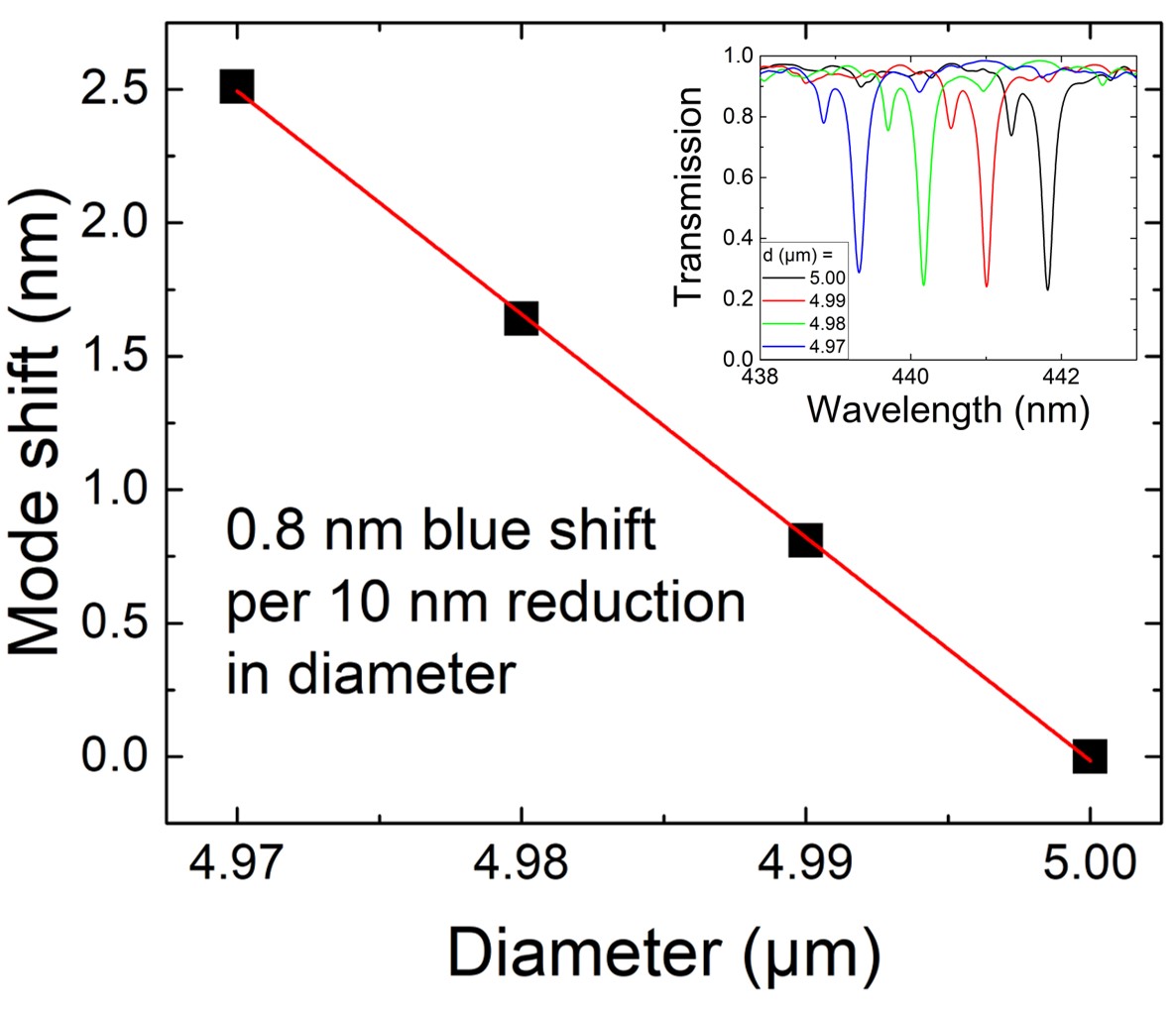}
  \caption{Mode shift over diameter. The linear fit gives a 0.8 nm blue shift per 10 nm reduction in diameter. The inset shows the FDTD transmission simulations of the 420 nm mode.}
  \label{fig:shift2}
\end{figure}

Three-dimensional (3D) finite-difference time-domain (FDTD) simulations of devices with 5 $\mu$m diameter were performed with a Gaussian source placed in one end of the waveguide for transmission simulations. Fig. \ref{fig:spec} shows the simulated transmitted radiative flux of a $5~\mu \text{m}$ diameter microdisk, $A=90^\circ$, $w= 170$ nm, and $g=50$ nm. The emitted flux is measured at the end of the waveguide and divided by the flux at the beginning of the waveguide to obtain the transmission. Refractive indices at 7 different wavelengths are used to take dispersion into account. The simulation is compared to the experimental CW spectrum (bottom) of Fig. 2(a) in the main text with $g=45$ nm. The positions of first-order radial modes match well between simulation and experiment and the mode spacing is 3.6 to 4.0 nm. The theoretical mode spacing $\Delta \lambda$ is given by
\begin{equation}
    \Delta \lambda = \frac{\lambda^2}{2\pi r n_g},
\end{equation}
where $n_g$ is the group index. For bulk AlN $n_g=2.33$ and for bulk GaN $n_g=3.23$ at $\lambda= 430$ nm. This gives an effective $n_{g,eff}= 3.2$ for the investigated microresonator and an expected mode spacing of $\Delta \lambda = 3.7$ nm, matching very well with the experimental and simulated results. The azimuthal orders of the modes are identified in Fig. \ref{fig:spec} and are determined by counting the nodes in the $H_z$ field. Modes of orders $m=85$ to 79 are visible.

The simulated $H_z$ field of devices with $5~\mu \text{m}$ diameter and $g=30$ to 50 nm are depicted in Fig. \ref{fig:sim5}. Good phase matching can be observed for $g=40$ and 50 nm. The depicted mode is a first order radial mode with azimuthal order $m=79$ at 441 nm.  

The device geometry in the simulation is slightly different than for the fabricated devices. However, the region of interest, the coupling region is the same. The waveguide bends around the disk in a $90^\circ$ angle at a distance $g$ from the disk.
The only thing missing in the simulation is the waveguide bending away from the disk in a region that is irrelevant for coupling, which will only provide some minor bending losses that we can neglect in our simulation as we are focusing on the coupling.

Figure \ref{fig:shift2} shows the mode shift over diameter, indicating that per 10 nm reduction in diameter the mode blue shifts by 0.8 nm. The inset shows the corresponding transmission simulations for a mode at 441 nm. This phenomenon is observed experimentally due to the proximity effect of the waveguide with decreasing gap during the fabrication process and is discussed in Fig. 6 in the main text.

\begin{figure}[htbp]
\centering
\includegraphics[width=0.6\linewidth]{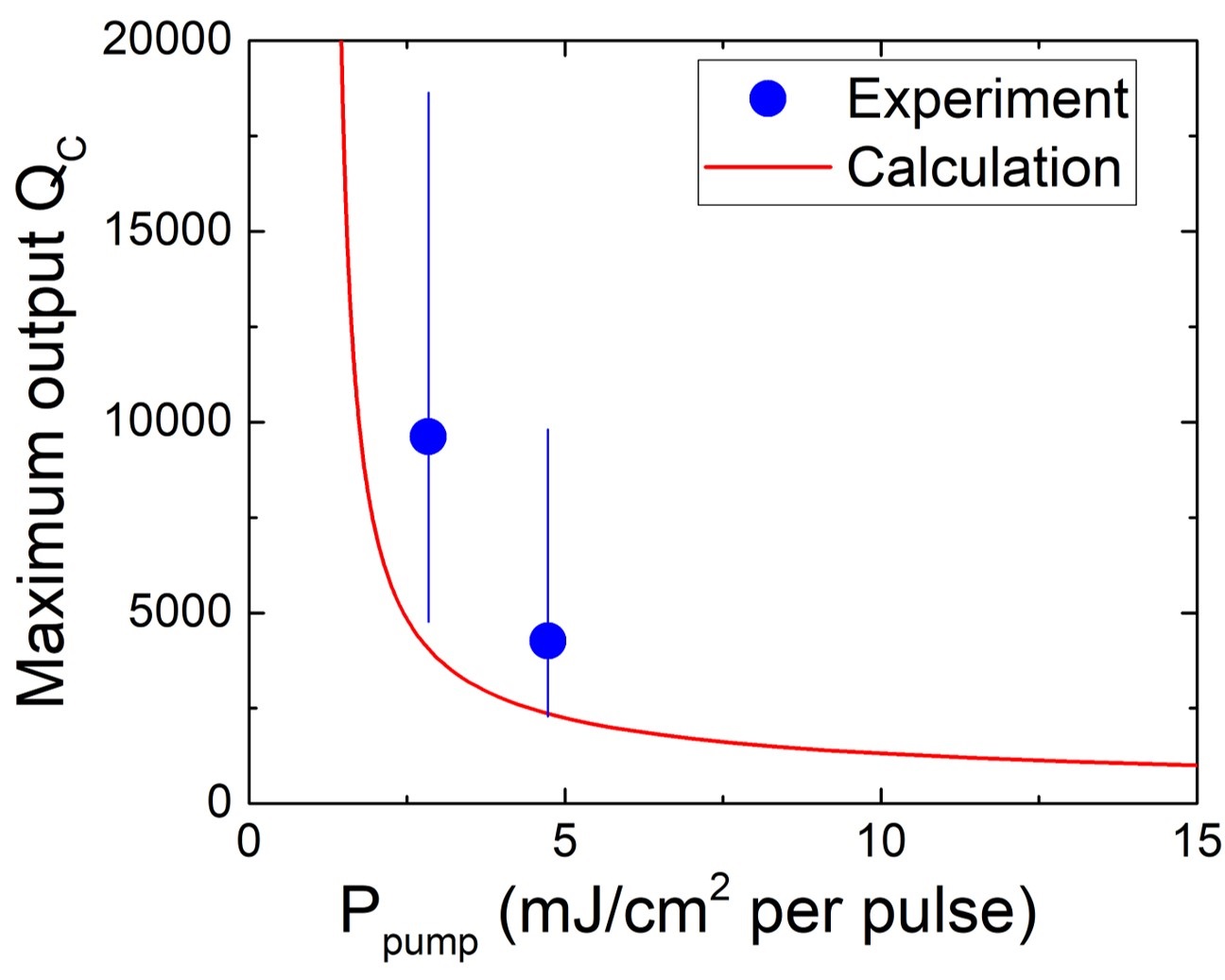}
  \caption{Maximum output $Q_C$ as a function of $P_{\text{pump}}$ according to equation (\ref{eq:Ppump}) and with the experimental values from Fig. 5 in the main text with error bars.}
  \label{fig:Qc}
\end{figure}

\begin{figure}[htbp]
\centering
\includegraphics[width=0.8\linewidth]{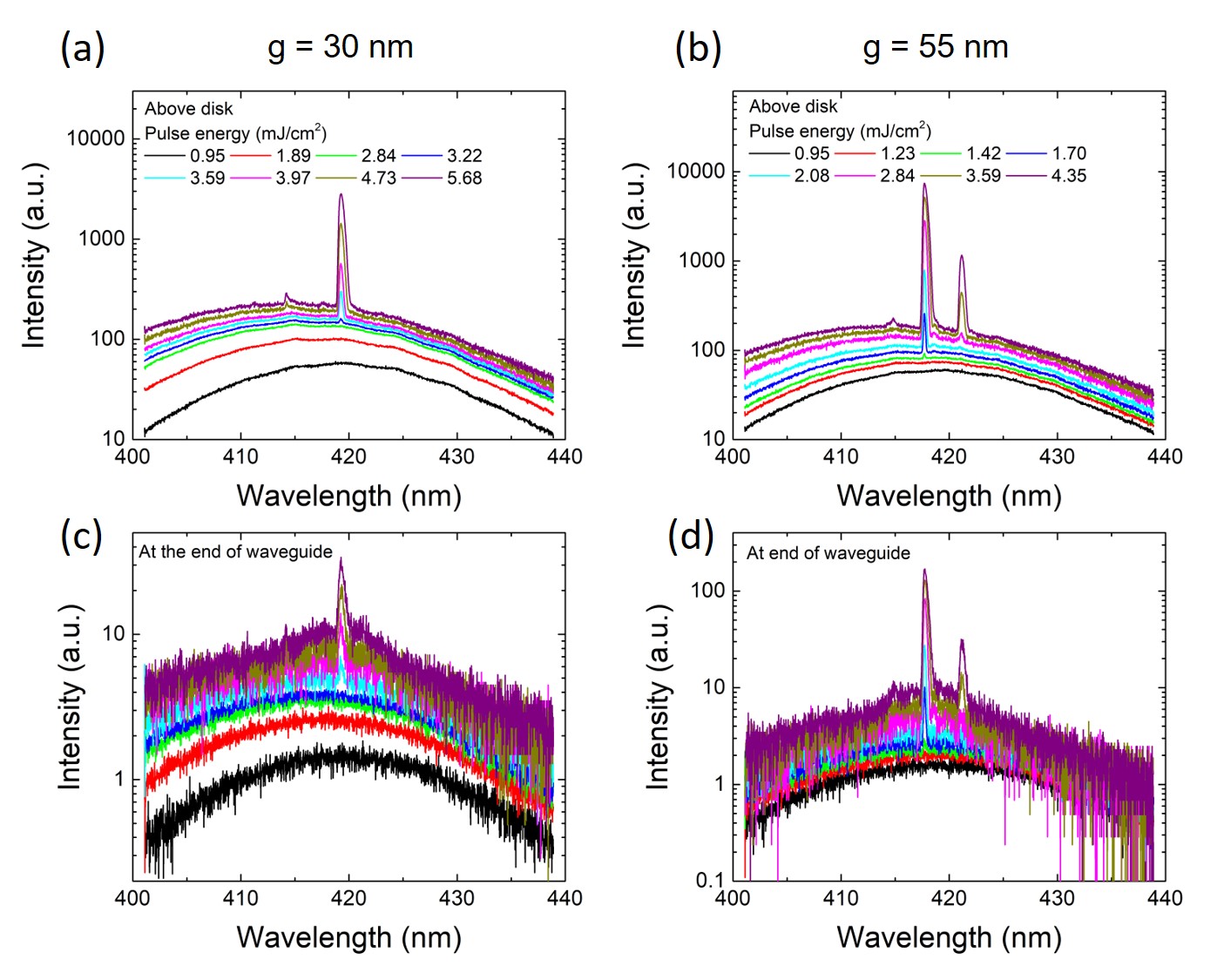}
  \caption{Lasing spectra measured above the disk for (a) $g= 30$ nm and (b) $g= 55$ nm and at the end of the waveguide for (c) $g= 30$ nm and (d) $g= 55$ nm.}
  \label{fig:3050}
\end{figure}

\section*{Maximum output power}

In order to determine the $Q_C$ value where $P_{\text{out}}$ becomes maximal for a given $P_{\text{pump}}$ we calculate $\text{d}P_{\text{out}}/\text{d}Q_C=0$, which gives

\begin{equation}
P_{\text{pump}} = F + \frac{BC}{Q_C^2}\cdot \frac{(Q_{\text{int}}+Q_C)^2}{C Q_{\text{int}}}.\label{eq:Ppump}
\end{equation}

Figure \ref{fig:Qc} shows the resulting maximum output $Q_C$ over $P_{\text{pump}}$. In the experimentally investigated range of 2 to 5 mJ/cm$^2$ per pulse maximum $P_{\text{out}}$ is achieved for $Q_C$ values in the range of $7000$ to $2000$. The experimental values are taken from Fig. 5 in the main text. There are many parameters that go into the fit in Fig. \ref{fig:Qc}, such as the fit of $P_{th}$ over $Q_{\text{loaded}}$ (Fig. 4 (d)), the fit of $P_{out}$ over $Q_C$ (Fig. 5), and $Q_{int}$, all of which have an uncertainty. The here shown fit is only intended to show a tendency based on the different parameters extracted from Fig. 4 and 5 and modelling.

\section*{Lasing and mode shift}

\begin{figure}[htbp]
\centering
\includegraphics[width=0.8\linewidth]{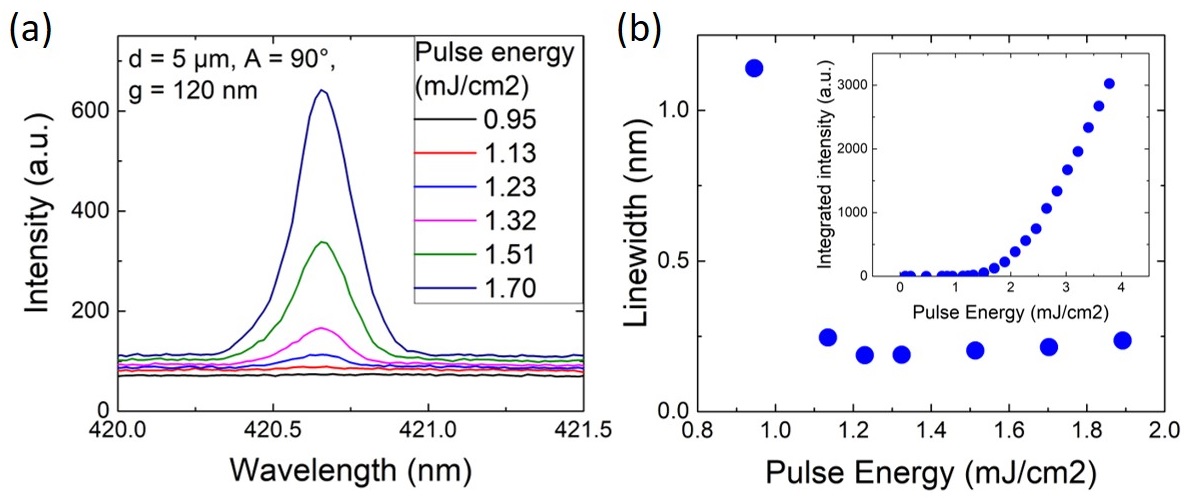}
  \caption{(a) Pulse-energy dependent spectra around threshold for a device with $A=90^\circ$ and $g=120$ nm. The threshold is at $1.2~\text{mJ/cm}^2$ per pulse. (b) Linewidth vs. pulse energy of the mode in (a) and (inset) integrated intensity vs. pulse energy for the same mode.}
  \label{fig:thresh2}
\end{figure}

\begin{figure}[htbp]
\centering
\includegraphics[width=0.8\linewidth]{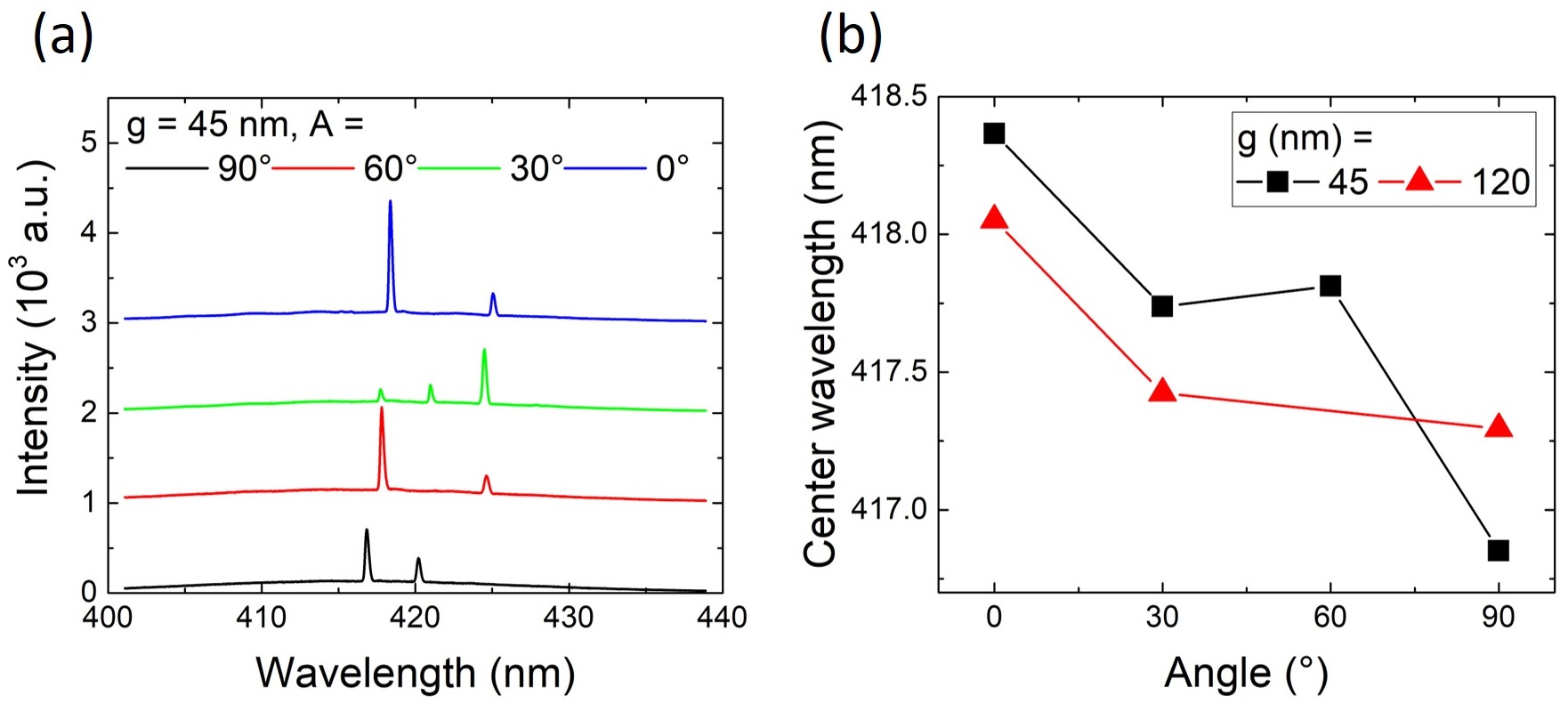}
  \caption{(a) Pulsed lasing spectra measured at the disk for devices with $g = 45 \text{~nm}$ for different angles. (b) Peak wavelength as a function of angle for $g = 45 \text{~nm}$ and $g = 120 \text{~nm}$.}
  \label{fig:lase2}
\end{figure}

Pulse-energy density dependent measurements of devices with $g= 30$ nm and $g= 55$ nm are shown in Fig. \ref{fig:3050} measured both above the disk and at the end of the waveguide. It can be clearly seen that the threshold decreases with increasing gap, as is discussed in the main text in respect to Fig. 4.

Figure \ref{fig:thresh2} (a) shows pulse-energy dependent spectra around the threshold for a device with $A=90^\circ$ and $g=120$ nm. The threshold is determined to be $1.2~\text{mJ/cm}^2$ per pulse, which corresponds to $300~\text{kW/cm}^2$. The linewidth over pulse energy is shown Fig. \ref{fig:thresh2} (b), indicating a linewidth narrowing of more than a factor of 2. The inset in  Fig. \ref{fig:thresh2} (b) shows the integrated intensity over pulse energy.

Figure \ref{fig:lase2} (a) shows spectra for devices with $g=45$ nm and angles from $A= 0^\circ$ to $90^\circ$. Figure \ref{fig:lase2} (b) shows the peak wavelength vs. $A$ for $g=45$ and 120 nm. For small angles a constant 0.3 nm blue shift is observed when going from $g=45$ to 120 nm, due to the change in effective refractive index. For small angles the diameter reduction due to waveguide proximity is negligible. At $A= 90^\circ$ a 0.4 nm red shift is observed when going from $g=45$ to 120 nm due to the increase in disk diameter of approximately 5 nm, given by Fig. \ref{fig:shift2}. With increasing angle a blue shift is observed due to a reduction in disk diameter, which is more pronounced at a smaller gap size. The blue shifts of 2 nm at $g=45$ nm and 1 nm at $g=120$ nm correspond to 20 nm and $10 ~\text{nm}$ reduction in disk diameter, respectively.

\end{document}